\documentclass[journal=jctcce,manuscript=article]{achemso}
\setkeys{acs}{maxauthors=10,etalmode=truncate,chaptertitle =true,articletitle=true}
\newcommand{\onlinecite}[1]{\hspace{-1 ex} \nocite{#1}\citenum{#1}} 
\usepackage{amssymb}
\usepackage{amsmath}
\usepackage[usenames,dvipsnames]{xcolor}
\usepackage{graphicx}
\usepackage{float}
\usepackage[normalem]{ulem}
\usepackage{subcaption}
\usepackage{epstopdf}
\usepackage{natbib}

\newcommand{\ket}[1]{\left|#1\right\rangle}

\newcommand{\abs}[1]{\left|#1\right|}

\newcommand{\qexpt}[3]{\left\langle#1 \left| #2 \right| #3\right\rangle}

\usepackage{xcolor}

\usepackage{multirow}
\usepackage{bm}

\title{Excited state orbital optimization via minimizing the square of the gradient: General approach and application to singly and doubly excited states via density functional theory.}

\author{Diptarka Hait}
\email{diptarka@berkeley.edu}
\affiliation
{{Kenneth S. Pitzer Center for Theoretical Chemistry, Department of Chemistry, University of California, Berkeley, California 94720, USA}}
\alsoaffiliation{Chemical Sciences Division, Lawrence Berkeley National Laboratory, Berkeley, California 94720, USA}
\author{Martin Head-Gordon}
\email{mhg@cchem.berkeley.edu}
\affiliation
{{Kenneth S. Pitzer Center for Theoretical Chemistry, Department of Chemistry, University of California, Berkeley, California 94720, USA}}
\alsoaffiliation{Chemical Sciences Division, Lawrence Berkeley National Laboratory, Berkeley, California 94720, USA}

\begin{document}
	\begin{abstract}
    We present a general approach to converge excited state solutions to \textit{any} quantum chemistry orbital optimization process, without the risk of variational collapse. The resulting Square Gradient Minimization (SGM) approach \textcolor{black}{only requires analytic energy/Lagrangian orbital gradients and} merely costs 3 times as much as ground state orbital optimization (per iteration), when implemented via a finite difference approach
    .  SGM is applied to both single determinant $\Delta$SCF and spin-purified Restricted Open-Shell Kohn-Sham (ROKS) approaches to study the accuracy of orbital optimized DFT excited states. It is found that SGM can converge challenging states where the Maximum Overlap Method (MOM) or analogues either collapse to the ground state or fail to converge. We also report that $\Delta$SCF/ROKS predict highly accurate excitation energies for doubly excited states (which are inaccessible via TDDFT). Singly excited states obtained via ROKS are also found to be quite accurate, especially for Rydberg states that frustrate (semi)local TDDFT. Our results suggest that orbital optimized excited state DFT methods can be used to push past the limitations of TDDFT to doubly excited, charge-transfer or Rydberg states, making them a useful tool for the practical quantum chemist's toolbox for studying excited states in large systems.
	\end{abstract}
	\maketitle
	\section{Introduction}
Accurate quantum chemical methods for modeling electronic excited states are essential for gaining insight into the photophysics and photochemistry of molecules and materials. The most widely used technique for excited state calculations at present is time dependent density functional theory (TDDFT)\cite{runge1984density,casida1995time,bauernschmitt1996treatment,marques2004time,dreuw2005single}, on account of its relatively low computational complexity ($O(N^{2-3})$, where $N$ is the molecule size) and reasonable accuracy for many problems\cite{furche2002adiabatic,jacquemin2009extensive}. TDDFT excited states are computed via determining the linear response of a ground state DFT solution to time-dependent external electric fields\cite{dreuw2005single}, permitting simultaneous modeling of multiple excited states. In principle, TDDFT is formally exact \cite{runge1984density} when the exact exchange-correlation (xc) ground state functional is employed, although lack of that functional, and the need for the widely used adiabatic local density approximation\cite{bauernschmitt1996treatment,marques2004time,dreuw2005single} (ALDA) prevents this from being the case in practice. ALDA in fact restricts utility of TDDFT to single excitations out of the reference alone, with large errors arising whenever the target excited state has significant double (or higher) excitation character\cite{tozer2000determination,maitra2004double,levine2006conical,elliott2011perspectives}. Furthermore, TDDFT is known to systematically underestimate excitation energies for charge-transfer \cite{dreuw2003long,dreuw2004failure,dreuw2005single,peach2008excitation} and Rydberg\cite{casida1998molecular,tozer2000determination} states, and yields qualitatively erroneous potential energy surfaces along single bond dissociation coordinates\cite{hait2019beyond}. These effects originally stem from errors in the ground state DFT solution like delocalization error\cite{perdew1982density,hait2018delocalization} or spin symmetry breaking\cite{szabo2012modern,hait2019wellbehaved}, but the linear response protocol augments these deficiencies in the reference to catastrophic levels in excited states, on account of insufficient orbital relaxation \cite{ziegler2008revised,subotnik2011communication,zhekova2014perspective,hait2016prediction}. 

Orbital-optimized excited state methods have consequently seen a renewal of interest in recent years\cite{gilbert2008self,thom2008locating,liu2012communication,kowalczyk2013excitation,krykunov2014introducing,zhao2016efficient,shea2018communication,ye2017sigma,ye2019half,tran2019tracking,nite2019low,burton2019general,zhao2019variational,shea2019generalized}, and have been successfully applied to problems like core excitations\cite{besley2009self,oosterbaan2018non} and CT states\cite{hait2016prediction,liu2012communication} where orbital relaxation is expected to play a key role. However, excited state orbital optimization is fundamentally a challenging task due to the possibility of collapsing back into the ground state (often described as ``variational collapse") as excited states are typically saddle point solutions of the orbital optimization equations. The Maximum Overlap Method (MOM)\cite{gilbert2008self} attempts to mitigate this by selecting occupied orbitals after each iteration via maximization of overlap with the occupied orbitals from the previous iteration (instead of filling orbitals in ascending order of their energies). MOM has been quite successful in reducing the frequency of variational collapse, but has not fully eliminated it in practice\cite{mewes2014molecular,barca2018simple}.

The continuing spectre of variational collapse has subsequently led to attempts to develop alternative variational principles\cite{zhao2016global,ye2017sigma,shea2018communication} where excited states are true minima instead of saddle points. Such principles often employ the energy variance, which involves the $\hat{\mathbf{H}}^2$ operator. The matrix elements of $\hat{\mathbf{H}}^2$ are quite computationally challenging due to the presence of three and four particle operators. Furthermore, a straightforward generalization to DFT is not possible as xc functionals approximate $\langle \hat{\mathbf{H}}\rangle$ and not $\langle \hat{\mathbf{H}}^2\rangle$, making it difficult to capitalize on the enormous strides made in Kohn-Sham DFT (KS-DFT) functional development for (ground state) energetics\cite{goerigk2017look,mardirossian2017thirty,najibi2018nonlocal,mardirossian2018survival} and properties\cite{hait2018accurate,hait2018accuratepolar} in recent years. 

In this work, we present a general approach that can extend \textit{any} ground state orbital-optimization method to excited states, without any apparent onset of variational collapse. The computational cost of this approach is only about 3 times the cost of ground state orbital optimization (per iteration), when  implemented via a simple finite-difference protocol based on analytic orbital gradients of the energy/Lagrangian. This method is subsequently applied to two excited state orbital optimized DFT techniques: $\Delta$SCF and Restricted Open-Shell Kohn-Sham (ROKS). The utility of these approaches is demonstrated via application to theoretically well characterized double excitations of small molecules, singly excited states of formaldehyde and an analysis of the absorption spectrum of zinc phthalocyanine. 

\section{Theory}
\subsection{Objective Function}
Orbital optimization based methods attempt to minimize some Lagrangian $\mathcal{L}$ against orbital degrees of freedom $\vec{\theta}$ (that mix occupied orbitals $\{i\}$ with virtual orbitals $\{a\}$). $\mathcal{L}$ is simply the energy for Hartree-Fock (HF) or DFT (or indeed, any variational method), although it can be considerably more complex (eg. including amplitude constraint terms) for non-variational methods like M{\"o}ller-Plesset perturbation theory\cite{lochan2007orbital,neese2009assessment,bozkaya2014oomp2,lee2018regularized} or coupled cluster.\cite{sherrill1998energies} For excited states, the typical objective is to find an unstable extremum of $\mathcal{L}$ instead of the global minimum, which is quite challenging due to the possibility of variational collapse down to a minimum. 

We convert the extremization problem into a minimization by instead focusing on:
\begin{align}
    \Delta = \abs{\nabla_{\vec{\theta}} \mathcal{L}}^2=\displaystyle\sum\limits_{ai}\abs{\dfrac{\partial \mathcal{L}}{\partial \theta_{ai}}}^2
\end{align}
$\Delta$ therefore is merely the square of the gradient of $\mathcal{L}$ against orbital degrees of freedom $\vec{\theta}$, and is therefore positive semidefinite by construction. $\Delta=0$ if and only if $\nabla_{\vec{\theta}} \mathcal{L}=0$, which  indicates stationarity of $\mathcal{L}$ against the orbital degrees of freedom. The challenges typically encountered in optimizing unstable extrema (i.e. saddle points or maxima) in $\mathcal{L}$ are therefore averted, as every orbital optimized state is a global minima of $\Delta$. Other extrema are possible, as discussed later, but are easily identifiable by $\Delta\ne 0$.

\subsection{Gradient}
The gradient of $\Delta$ with respect to $\vec{\theta}$ is given by: 
\begin{align}
    \dfrac{\partial}{\partial \theta_{ai}}\Delta = \dfrac{\partial}{\partial \theta_{ai}}\displaystyle\sum\limits_{bj}\abs{\dfrac{\partial \mathcal{L}}{\partial \theta_{bj}}}^2=\displaystyle\sum\limits_{bj}\left(\dfrac{\partial \mathcal{L}}{\partial \theta_{bj}}\right)^*\left(\dfrac{\partial^2 \mathcal{L}}{\partial \theta_{bj}\partial \theta_{ai}}\right)+h.c.\label{deltadir}
\end{align}
For HF/DFT, the cost of analytically evaluating the gradient via the matrix-vector contraction $\displaystyle\sum\limits_{bj}\left(\dfrac{\partial \mathcal{L}}{\partial \theta_{bj}}\right)^*\left(\dfrac{\partial^2 \mathcal{L}}{\partial \theta_{bj}\partial \theta_{ai}}\right)$ should roughly equal the cost of constructing the Fock matrix $\hat{\mathbf{F}}$. The cost of analytically evaluating $\nabla_{\vec{\theta}} \Delta$ should therefore be twice the cost of evaluating $\nabla_{\vec{\theta}} \mathcal{L}$: once for constructing $\nabla_{\vec{\theta}} \mathcal{L}$, and another for the contraction with the Hessian $\left(\dfrac{\partial^2 \mathcal{L}}{\partial \vec{\theta}\partial  \vec{\theta^\prime}}\right)$. Analytical $\Delta$ gradients are therefore straightforward at the HF/DFT level, at a compute cost of roughly twice the analytical $\mathcal{L}$ orbital gradient. However, efficient implementation of the analytic $\mathcal{L}$ Hessian is undoubtedly challenging for more complex methods. 

A simple finite difference approach permits us to sidestep this issue for very little additional cost. Such an approach has already been used for orbital stability analysis (i.e. extremal eigenvalues of the orbital Hessian)\cite{sharada2015wavefunction,lee2018regularized}, building on earlier work evaluating extremal eigenvalues of the force constant matrix (Hessian with respect to nuclear displacements).\cite{degelmann2002hess,reiher2003modes,kaledin2005modes,sharada2014hessian} We know that:
\begin{align}
\left(\dfrac{\partial \mathcal{L}}{\partial \theta_{ai}}\right)_{\vec{\theta}=\vec{\theta_0}+\vec{\delta \theta}}&=\left(\dfrac{\partial \mathcal{L}}{\partial \theta_{ai}}\right)_{\vec{\theta}=\vec{\theta_0}}+\displaystyle\sum\limits_{bj}\left(\dfrac{\partial^2 \mathcal{L}}{\partial \theta_{bj}\partial \theta_{ai}}\right)_{\vec{\theta}=\vec{\theta_0}}{\delta \theta}_{bj}+O((\vec{\delta\theta})^2)
\end{align}
from a Taylor expansion of the derivative $\dfrac{\partial \mathcal{L}}{\partial \theta_{ai}}$ about the point $\vec{\theta}=\vec{\theta_0}$, on account of a perturbation $\vec{\delta\theta}$. Subsequently, we can choose $\vec{\delta\theta}=\lambda \left(\nabla_{\vec{\theta}} \mathcal{L}\right)^*_{\vec{\theta}=\vec{\theta_0}}$, which yields:
\begin{align}
    &\displaystyle\sum\limits_{bj}\left(\dfrac{\partial \mathcal{L}}{\partial \theta_{bj}}\right)_{\vec{\theta}=\vec{\theta_0}}^*\left(\dfrac{\partial^2 \mathcal{L}}{\partial \theta_{bj}\partial \theta_{ai}}\right)_{\vec{\theta}=\vec{\theta_0}}=\dfrac{1}{2\lambda}\left(\left(\dfrac{\partial \mathcal{L}}{\partial \theta_{ai}}\right)_{\vec{\theta}=\vec{\theta_0}+\vec{\delta \theta}}-\left(\dfrac{\partial \mathcal{L}}{\partial \theta_{ai}}\right)_{\vec{\theta}=\vec{\theta_0}-\vec{\delta \theta}}\right)+O\left(\lambda^2\left( \left(\nabla_{\vec{\theta}} \mathcal{L}\right)^*_{\vec{\theta}=\vec{\theta_0}}\right)^3\right)\\
    &\implies \left(\nabla_{\vec{\theta}} \Delta\right)_{\vec{\theta}=\vec{\theta_0}}=\dfrac{1}{2\lambda}\left(\left(\nabla_{\vec{\theta}} \mathcal{L}\right)_{\vec{\theta}=\vec{\theta_0}+\vec{\delta \theta}}-\left(\nabla_{\vec{\theta}} \mathcal{L}\right)_{\vec{\theta}=\vec{\theta_0}-\vec{\delta \theta}}\right)+O\left(\lambda^2\left( \left(\nabla_{\vec{\theta}} \mathcal{L}\right)^*_{\vec{\theta}=\vec{\theta_0}}\right)^3\right)
\end{align}
In other words, taking the finite difference between the gradient $\nabla_{\vec{\theta}} \mathcal{L}$ at two slightly shifted $\vec{\theta}$ (with the shift being proportional to the gradient $\nabla_{\vec{\theta}} \mathcal{L}$ at the central point) yields the desired Hessian-gradient contraction. The cost of this approach for finding $\nabla_{\vec{\theta}} \Delta$ is therefore thrice the cost of a single $\nabla_{\vec{\theta}} \mathcal{L}$ gradient evaluation, which is not a substantial increase over the $2\times$ cost associated with contraction with the analytic Hessian. While this approach does introduce precision errors associated with finite differencing, their magnitude can be controlled via judicious choice of $\lambda$. More importantly, the errors scale as $O\left(\lambda^2\left(\left(\nabla_{\vec{\theta}} \mathcal{L}\right)^*\right)^3\right)$, indicating that they are the largest when we are far from convergence (i.e. large $\nabla_{\vec{\theta}} \mathcal{L}$) when a very accurate gradient is not critical. The errors should be quite small close to convergence (when $\nabla_{\vec{\theta}} \mathcal{L}$ should be small). Alternative higher order finite difference formulae could also be employed, though we shall not consider such choices here. 
\subsection{Preconditioner}
The convergence of a gradient based optimization process can be dramatically accelerated via use of appropriate preconditioners, like a diagonal approximation to the Hessian. Unfortunately, exact evaluation of the diagonal terms of the $\Delta$ Hessian is likely far too computationally demanding to be worthwhile. 
However, mean-field terms (i.e. $\hat{\mathbf{F}}$ terms) make up the largest portion of $\mathcal{L}$ for non-strongly correlated species. Focusing on those terms alone suggests that within a pseudocanonical orbital basis (i.e. occupied-occupied and virtual-virtual blocks of $\hat{\mathbf{F}}$ are diagonal), an approximate preconditioner $B_{ia,jb}=8\left(\epsilon_a-\epsilon_i\right)^2\delta_{ia}\delta_{jb}$ (where $\epsilon_a$ and $\epsilon_i$ are energies of pseudocanonical spin-orbitals $a$ and $i$, respectively) would be appropriate. This is basically a generalization of the preconditioner used in the geometric direct minimization (GDM) method\cite{van2002geometric} for ground state minimization.

\subsection{The SGM method for orbital optimization}
The gradient and the preconditioner described in the two preceding subsections can be employed to minimize $\Delta$, starting from any initial guess in orbital space. To do so, we build upon the GDM quasi-Newton method\cite{fletcher2013practical} for $\mathcal{L}$ minimization (which uses the BFGS update\cite{broyden1970convergence,fletcher1970new,goldfarb1970family,shanno1970conditioning}). For our squared gradient minimization (SGM) problem, we supply the gradient and preconditioner appropriate for $\Delta$ to the GDM algorithm. The computational cost of a single SGM iteration should therefore be at most three times the cost of the corresponding GDM iteration for the ground state.

SGM would ideally converge to the closest solution in orbital space, when supplied with an initial set of guess orbitals. In practice however, the highly approximate nature of the initial preconditioner could result in large initial steps that lead to convergence to an alternative root (or even the ground state!). However, this can be easily mitigated by scaling the gradient down by some scalar $c$ to minimize the size of the initial steps in order to prevent large initial stepsizes. The approximate BFGS Hessian would however be effectively scaled by the same $c$, and the long term convergence rate not be (too) negatively impacted. We have found that $c=1$ is typically adequate for most cases, but a very low value of $c=0.01$ could be employed as a conservative choice for difficult cases. 



\subsection{Relationship with other methods}

$\Delta$ minimization via SGM is essentially a generalization of GDM\cite{van2002geometric} for saddle point optimization. It is consequently a direct minimization based alternative to $\hat{\mathbf{F}}$ matrix diagonalization methods like the Maximum Overlap Methods (MOM)\cite{gilbert2008self,barca2018simple}. \textcolor{black}{In addition, SGM is closely related to the excited state variational principle employed by the Excited State Mean-Field (ESMF) approach described in Ref \onlinecite{shea2018communication} and the $\sigma$-SCF approach described in Ref \onlinecite{ye2017sigma}. The objective function in Ref \onlinecite{shea2018communication} reduces to $\Delta$ if its energy targeting component is deleted. This has very recently been generalized to a generalized variational principle (GVP)\cite{shea2019generalized}, which smoothly scales various components, and can thus become $\Delta$ in some limits. The presence of the energy targeting term enables GVP to target states close to a particular input energy, while SGM $\Delta$ minimization aims to converge to the closest minima in orbital space to the initial guess. SGM is therefore simpler, though necessarily more guess-dependent. It is also worth noting that a finite difference approach was employed in Ref \onlinecite{shea2019generalized} for Newton-Raphson iterations, although the per-iteration computational cost was larger due to the need to construct a large Krylov subspace for inverting the Hessian of the objective function.}

The parallels with $\sigma$-SCF\cite{ye2017sigma} are less obvious at first glance. $\sigma$-SCF minimizes the energy variance $\sigma^2=\qexpt{\Phi}{\left(\hat{\mathbf{H}}-\langle \hat{\mathbf{H}}\rangle\right)^2}{\Phi}=\qexpt{\Phi}{\hat{\mathbf{H}^2}}{\Phi}-\left(\qexpt{\Phi}{\hat{\mathbf{H}}}{\Phi}\right)^2$ for a single determinant $\ket{\Phi}$. The computational expense of evaluation of $\qexpt{\Phi}{\hat{\mathbf{H}^2}}{\Phi}$ makes it natural to wonder if substitution of $\hat{\mathbf{H}}$ with a mean-field 1 body Hamiltonian like the Fock-matrix $\hat{\mathbf{F}}$ would be acceptable. The most challenging term would then be:
\begin{align}
    \qexpt{\Phi}{\hat{\mathbf{F}^2}}{\Phi}&=\displaystyle\sum\limits_{\ket{D}}\qexpt{\Phi}{\hat{\mathbf{F}}}{D}\qexpt{D}{\hat{\mathbf{F}}}{\Phi}
\end{align}
by doing a resolution of the identity over all determinants $\ket{D}$ in Hilbert space. $\qexpt{\Phi}{\hat{\mathbf{F}}}{D}\ne 0$ only if $\ket{D}$ is either $\ket{\Phi}$ or a single excitation $\ket{\Phi^a_i}$. Therefore:
\begin{align}
    \qexpt{\Phi}{\hat{\mathbf{F}^2}}{\Phi}&=\left(\qexpt{\Phi}{\hat{\mathbf{F}}}{\Phi}\right)^2+\displaystyle\sum\limits_{i,a}\qexpt{\Phi}{\hat{\mathbf{F}}}{\Phi^a_i}\qexpt{\Phi^a_i}{\hat{\mathbf{F}}}{\Phi}\\
    \implies \qexpt{\Phi}{\hat{\mathbf{F}^2}}{\Phi}-\left(\qexpt{\Phi}{\hat{\mathbf{F}}}{\Phi}\right)^2&=\displaystyle\sum\limits_{i,a}\qexpt{\Phi}{\hat{\mathbf{F}}}{\Phi^a_i}\qexpt{\Phi^a_i}{\hat{\mathbf{F}}}{\Phi}=\displaystyle\sum\limits_{i,a}\abs{F_{ai}}^2=\dfrac{1}{4}\Delta
\end{align}
In essence, SGM (or any single determinant optimizer like MOM) performs mean-field variance minimization, in contrast to the full $\mathbf{\hat{H}}$ based variance minimization of $\sigma$-SCF. 
\subsection{Local extrema in $\Delta$}

From Eqn \ref{deltadir}, we can infer that $\nabla_{\vec{\theta}} \Delta =0$ implies either $\nabla_{\vec{\theta}} \mathcal{L}=0$ (indicating successful extremization) or that the gradient $\nabla_{\vec{\theta}} \mathcal{L}$ belongs to the null-space of the Hessian $\dfrac{\partial^2 \mathcal{L}}{\partial \vec{\theta}\partial \vec{\theta^\prime}}$. While cases with singular $\dfrac{\partial^2 \mathcal{L}}{\partial \vec{\theta}\partial \vec{\theta^\prime}}$ are known in quantum chemistry as Coulson-Fischer points\cite{coulson1949xxxiv}, such points are defined by zero gradients. Hence little or nothing is known about singular orbital hessians with \textit{nonzero} gradients. Interestingly, we have encountered a few such solutions over the course of our investigations, as indicated by $\Delta \ne 0$ at convergence. We were able to escape them via use of ``better" initial guesses, such as by providing converged local spin-density approximation (LSDA) excited state orbitals to a hybrid DFT calculation (instead of ground state hybrid DFT orbitals).  

\section{Orbital optimized 
excited state DFT}
\subsection{$\Delta$SCF}
$\Delta$SCF\cite{ziegler1977calculation,kowalczyk2011assessment} methods converge a single Slater determinant as an excited state solution to the HF/KS equations. The likelihood of variational collapse had long restricted the utility of $\Delta$SCF, but the development of MOM led to a revival of interest in the method\cite{gilbert2008self,besley2009self,kowalczyk2011assessment,barca2017excitation}. MOM nonetheless does not always succeed in averting variational collapse (as will be shown later), making it desirable to have alternative solvers for challenging cases.

Apart from convergence, other main concerns with $\Delta$SCF are twofold. The Hohenberg-Kohn theorem\cite{hohenberg1964inhomogeneous} does not formally hold for excited states\cite{gaudoin2004lack}, and it cannot be assumed that ground state functionals will be accurate for excited states. $\Delta$SCF is thus a pragmatic choice for modeling excited states in large systems, but will not be a foolproof solution. Nonetheless, practical studies have shown that quite high levels of accuracy can be obtained from $\Delta$SCF\cite{gilbert2008self,besley2009self,kowalczyk2011assessment,barca2017excitation,barca2018simple} for challenging problems that TDDFT fails to address properly, without compromising accuracy in TDDFT's ideal domain of applicability (valence excitations in closed shell species). Our results also demonstrate this point, as will be shown later. 

The second, closely related, challenge facing $\Delta$SCF is that unlike ground states, excited states cannot often be well approximated by a single Slater determinant. In particular, single excitations out of closed-shell molecules are intrinsically multiconfigurational, as both $\alpha$ and $\beta$ electrons are equally likely to be excited (leading to at least two configurations of equal weights). $\Delta$SCF within the $M_S=0$ subspace can only target one of the configurations and would therefore yield a heavily spin-contaminated (``mixed") determinant with $\langle S^2 \rangle \approx 1$ for even otherwise well-behaved single excitations. This is an issue for singlet excited states alone, as $M_S=\pm 1$ triplet states are typically well described by single determinants and thus $\Delta$SCF. The singlet energies can be approximated via approximate spin-purification\cite{yamaguchi1988spin}, if the only major spin-contaminant is the corresponding triplet. It would however be ideal to orbital optimize the spin-purified energy directly instead of optimizing the mixed and triplet configurations separately. The Restricted Open-Shell Kohn-Sham (ROKS) method achieves this for pure open shell singlet states, and is described in the next subsection. Other potentially more general alternatives like half-projection\cite{ye2019half} or the DFT generalization to ESMF\cite{zhao2019variational} appear to possess double counting errors, making them somewhat less appealing. 


\subsection{ROKS}
The Restricted Open-Shell Kohn Sham (ROKS) technique\cite{filatov1999spin,kowalczyk2013excitation} optimizes orbitals for spin-pure singly excited states by extremizing
\begin{align}
    \mathcal{L}_\textrm{ROKS} = 2E_M - E_T 
\end{align}
where $E_M$ is the energy of the mixed determinant and $E_T$ is the energy of the triplet within the $M_S=1$ subspace, using the same spin-restricted orbitals. This is reasonable for true open shell singlets, as the mixed state is half singlet and half triplet when RO orbitals are used. The same strategy could also be applied to double excitations where a single electron pair has been broken, such as the $^1B_{3g}$ state of tetrazine\cite{loos2019reference}. ROKS has been shown to be quite effective at predicting HOMO$\to$LUMO type excitations in organic molecules\cite{kowalczyk2013excitation} and is excellent for CT state energies in systems where TDDFT fails catastrophically \cite{hait2016prediction}. However, the implementation described in Ref \onlinecite{kowalczyk2013excitation} is restricted to the lowest excited singlet ($S_1$) state alone. SGM however permits application of ROKS to arbitrary excited singlet states without collapse back to the $S_1$ state, thus considerably generalizing its applicability to excited state calculations. Of course, ROKS is itself limited in applicability: ROKS can only describe transitions that are well represented as promotions from one spatial occupied orbital to one spatial virtual orbital after orbital optimization. Excitations that can only be represented by transitions between multiple orbital pairs that have no common orbitals are unlikely to be well described on account of their natively multiconfigurational nature. A rather well-known example of the latter are the $L_b$ dark states in polyaromatic compounds\cite{prlj2016low}.

\section{Applications}

\subsection{Comparison of SGM to MOM and IMOM}

\begin{figure}[htb!]
\begin{minipage}{0.48\textwidth}
    \centering
    \includegraphics[width=\linewidth]{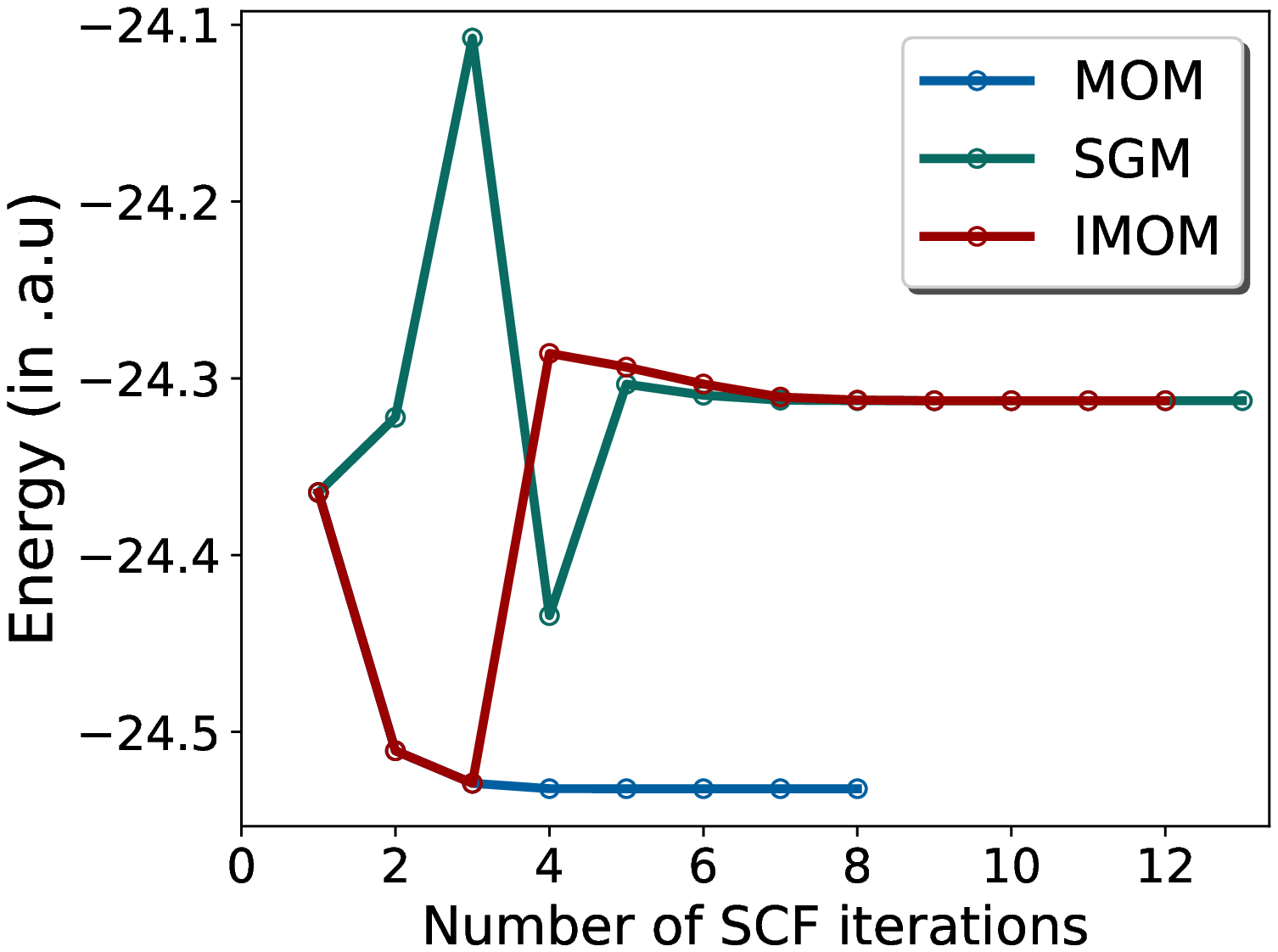}
    \label{fig:batomE}
\end{minipage}
\begin{minipage}{0.48\textwidth}
    \centering
    \includegraphics[width=\linewidth]{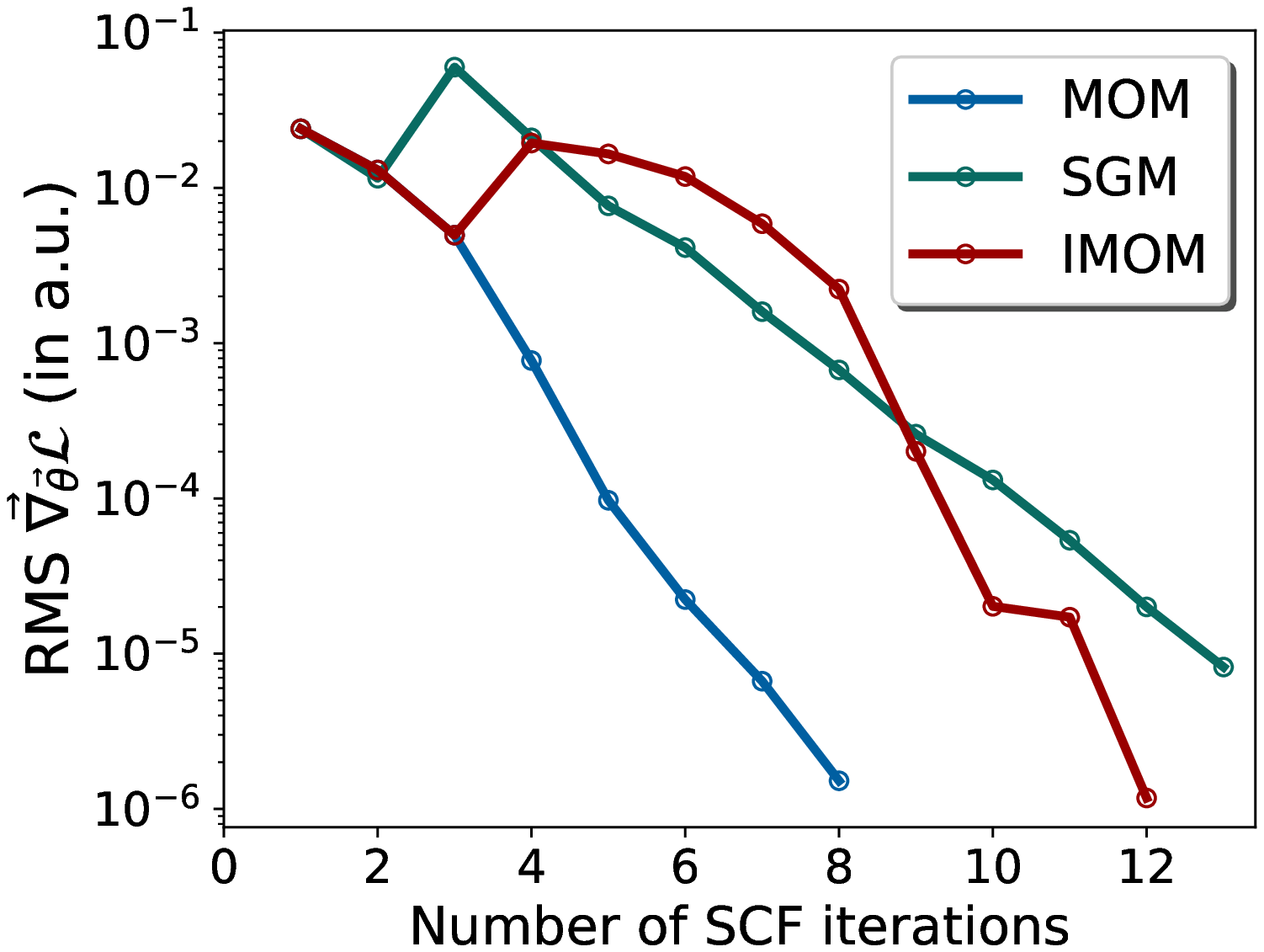}
    \label{fig:batomD}
\end{minipage}
\begin{minipage}{\textwidth}
    \vspace*{-15pt}
    \subcaption{2p$\to$3p of B atom (UHF/aug-cc-pVTZ\cite{dunning1989gaussian,kendall1992electron}).}
    \vspace*{-15pt}
    \label{fig:batomC}
\end{minipage}
\begin{minipage}{0.48\textwidth}
    \centering
    \includegraphics[width=\linewidth]{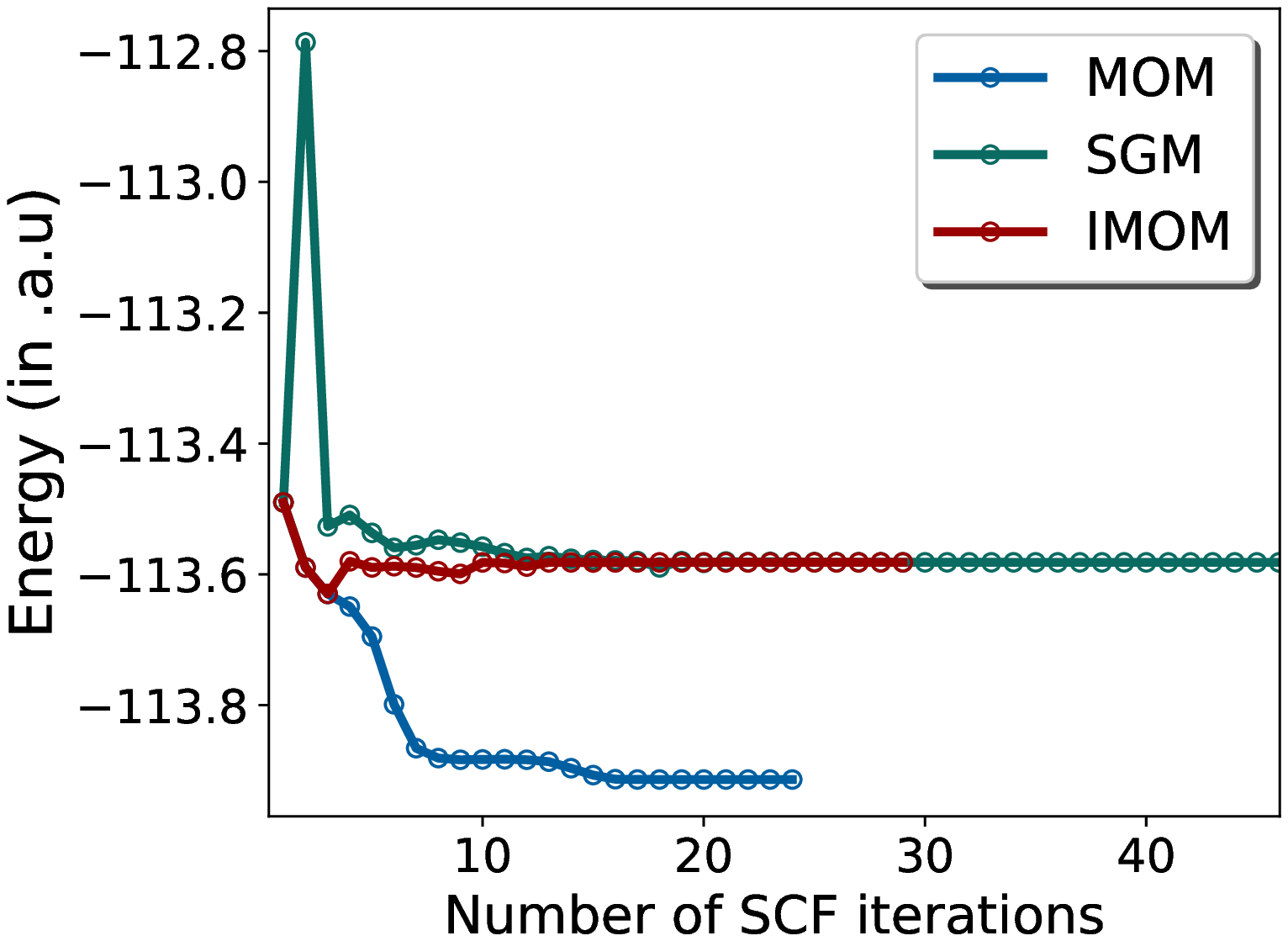}
    \label{fig:hchoE}
\end{minipage}
\begin{minipage}{0.48\textwidth}
    \centering
    \includegraphics[width=\linewidth]{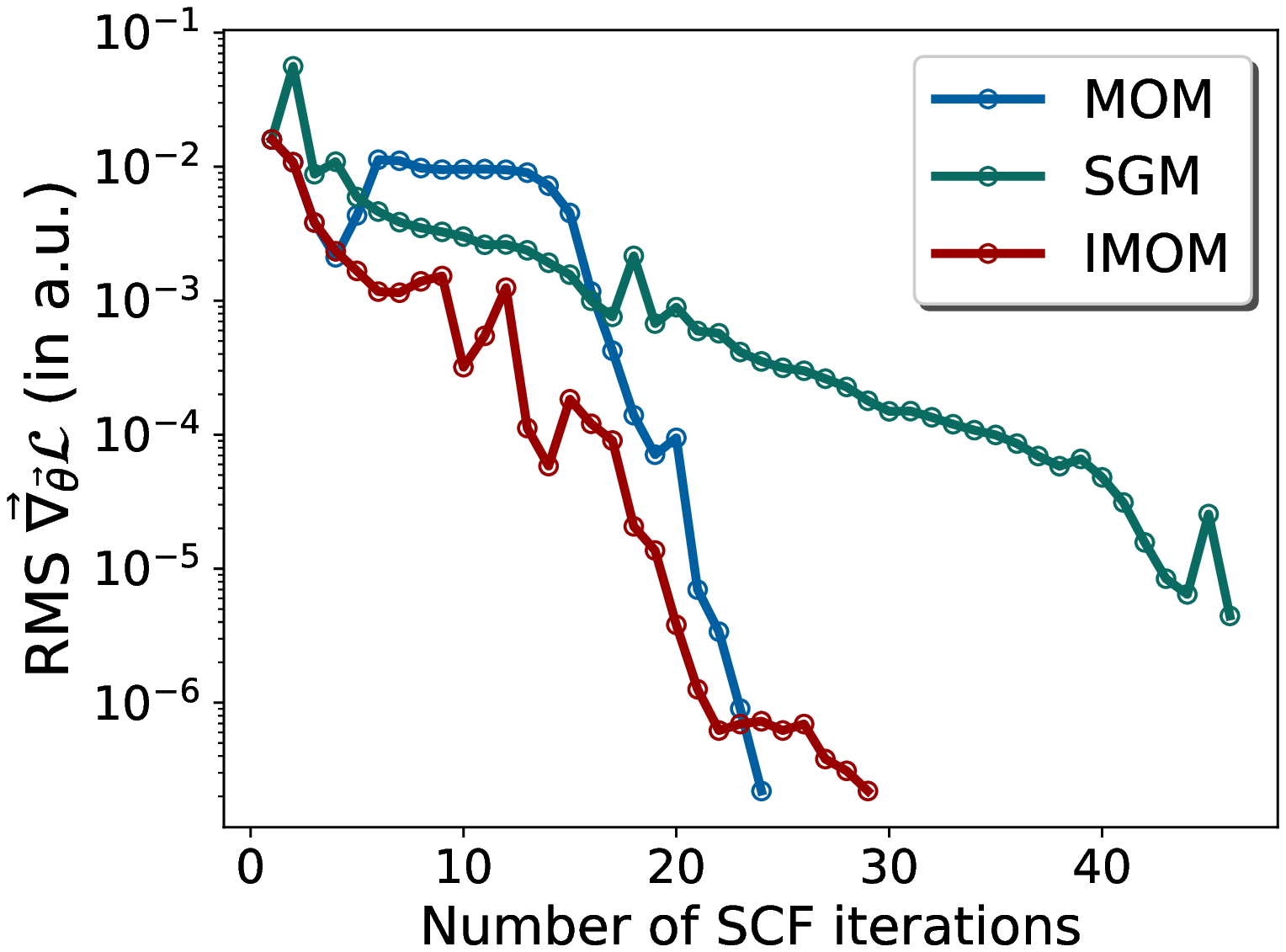}
    \label{fig:hchoD}
\end{minipage}
\begin{minipage}{\textwidth}
\vspace*{-15pt}
    \subcaption{O lone pair $\to$ C 4p$_y$ Rydberg state of HCHO (UHF/aug-cc-pVTZ\cite{dunning1989gaussian,kendall1992electron}).}
    \vspace*{-15pt}
    \label{fig:hchoC}
\end{minipage}
\caption{Energy and gradient ($\vec{\nabla}_{\vec{\theta}}\mathcal{L}$) convergence to $\Delta$SCF solutions with MOM, IMOM and SGM. 
SGM converges energies to $10^{-8}$ a.u. in 13 iterations (39 Fock builds) for the B atom and 46 iterations (138 Fock builds) for HCHO, with $c=1$. In contrast, IMOM requires 12 and 29 SCF cycles, respectively. 
}
\label{fig:momsgm}
\end{figure}

MOM has encountered considerable success in averting variational collapse for $\Delta$SCF, but is nonetheless not a perfect solution\cite{mewes2014molecular}. Two systems where MOM fails to avert variational collapse are the 2p$\to$ 3p excitation in the B atom\cite{barca2018simple} and a Rydberg-like single excitation out of the highest energy oxygen lone-pair to a C 4p$_y$ orbital in formaldehyde. SGM however is successful at converging both, as can be seen from the plots in Fig \ref{fig:momsgm}. As shown in Fig \ref{fig:momsgm}, the Initial MOM (IMOM) method\cite{barca2018simple} (which selects occupied orbitals at the end of a $\hat{\mathbf{F}}$ diagonalization based on overlap with an \textit{initial} set of orbitals vs the ones from the \textit{preceding} step) is also able to converge to the same solution as SGM for both of these cases, for a considerably smaller computational cost (stemming from fewer $\hat{\mathbf{F}}$ builds being required).

\begin{figure}[htb!]
\begin{minipage}{0.48\textwidth}
    \centering
    \includegraphics[width=\linewidth]{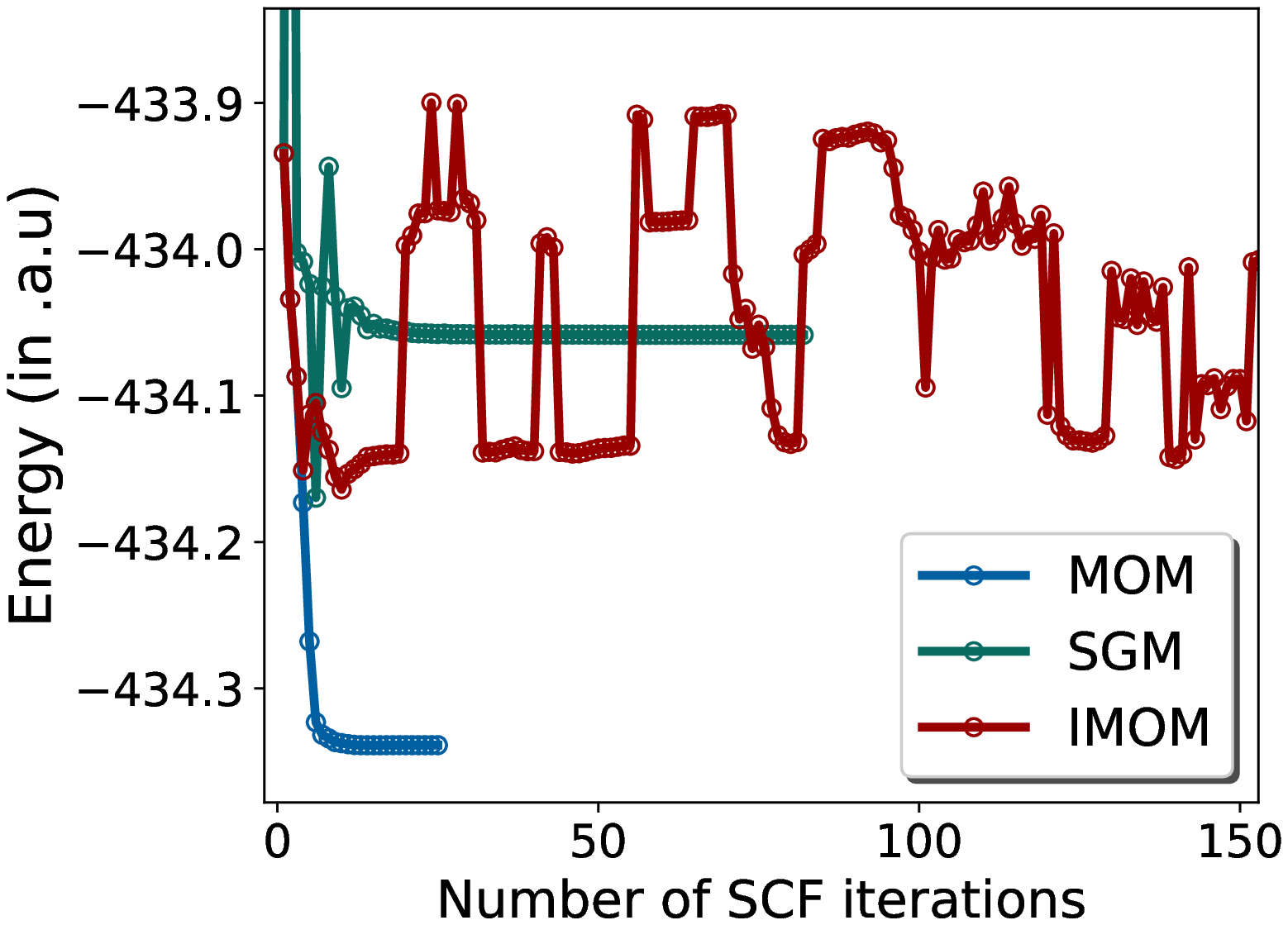}
    \label{fig:PhNO2E}
\end{minipage}
\begin{minipage}{0.48\textwidth}
    \centering
    \includegraphics[width=\linewidth]{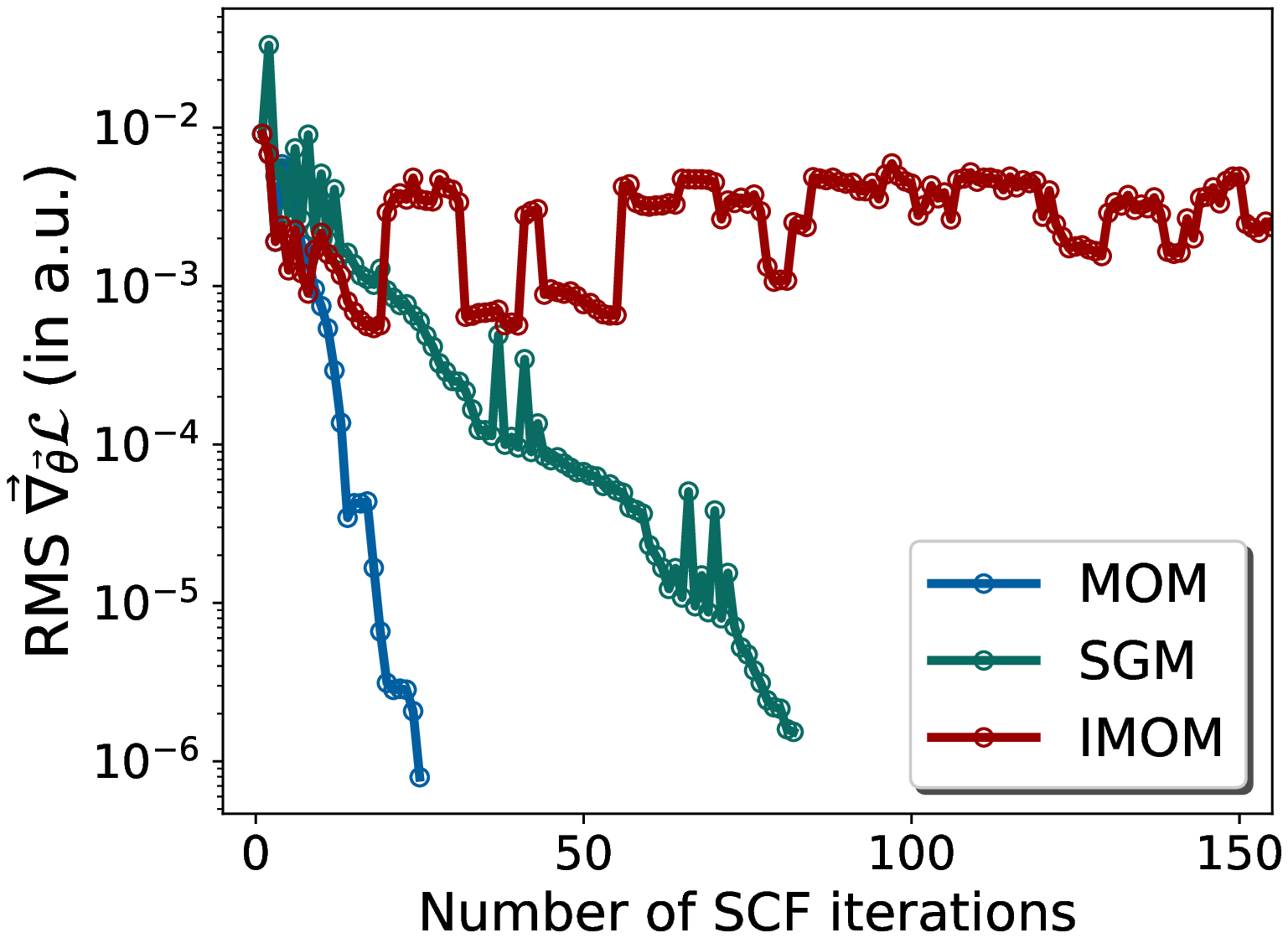}
    \label{fig:PhNO2D}
\end{minipage}
\caption{Energy and gradient ($\vec{\nabla}_{\vec{\theta}}\mathcal{L}$) convergence to $\Delta$SCF solutions for the excitation from the highest energy $\pi$ lone pair  to the second lowest $\pi^*$ orbital in nitrobenzene (UHF/def2-TZVP\cite{weigend2005balanced}). MOM collapses back to the ground state while IMOM fails to converge even after 500 iterations (only the first 150 are shown). SGM (with $c=1$) however converges the energy to $10^{-8}$ a.u. after 82 iterations (246 Fock builds).
}
\label{fig:imomsgm}
\end{figure}

IMOM's good performance stems from it avoiding `drifting' of orbitals away from the initial guess over multiple SCF iterations. However, IMOM can exhibit oscillatory behavior and fail to converge to a solution, as exhibited by the case of an excitation from the highest energy $\pi$ lone pair  to the second lowest $\pi^*$ orbital in nitrobenzene (as depicted in Fig \ref{fig:imomsgm}), where IMOM shows no sign of convergence even after 500 DIIS\cite{pulay1980convergence} steps.  This is likely a consequence of IMOM picking significantly different orbitals after some steps, based on discontinous ranking changes arising from small fluctuations in the overlap with the initial orbitals. On the other hand, MOM monotonically collapses back to the ground state. Considering all three examples, SGM appears to be a relatively stable $\Delta$SCF solver that could prove effective in converging challenging states, although it is more expensive than MOM or IMOM per iteration (and typically requires more steps when those methods converge).

\subsection{Application to Doubly Excited States}
Doubly excited states (or states with significant double excitation character) are typically inaccessible via TDDFT due to use of the ALDA. The efficacy of $\Delta$SCF for modeling such double excitations has already been hinted at \cite{barca2018simple,barca2017excitation}, leading us to study the extent to which ground state DFT functionals could reproduce vertical excitation energies for a few systems with theoretically well-characterized pure double excitations from  Ref \onlinecite{loos2019reference}. $\Delta$SCF solutions were examined for all states aside from the $^1B_{3g}$ state of tetrazine, where the presence of a broken electron pair necessitated use of ROKS (the corresponding doubly excited triplet was modelled with RO orbitals as well, to ensure spin-purity). 

\begin{table}[htb!]
\begin{tabular}{lrrrrrrrr}
\hline
  Species (Excitation)       & Rung 1 & Rung 2 & \multicolumn{2}{c}{Rung 3} & \multicolumn{2}{c}{Rung 4} & CC3   & TBE   \\ \hline
 & SPW92  & PBE    & SCAN        & B97M-V       & PBE0        & $\omega$B97X-V      &       &       \\
Be (2s$^2\to$ 2p$^2$)      & 6.97   & 6.98   & 7.11        & 7.08         & 7.23        & 7.52         & 7.08  & 7.06  \\
HNO (n$^2\to$ $\left(\pi^*\right)^2$)      & 4.00   & 4.13   & 4.24        & 4.33         & 4.24        & 4.26         & 5.21  & 4.32  \\
HCHO (n$^2\to$ $\left(\pi^*\right)^2$)    & 9.56   & 9.73   & 10.02       & 10.06        & 10.07       & 10.20        & 11.18 & 10.34 \\
C$_2$H$_4$ ($\pi^2\to$ $\left(\pi^*\right)^2$)    & 11.78  & 11.75  & 12.22       & 12.23        & 12.27       & 12.57        & 12.80 & 12.56 \\
CH$_3$NO (n$^2\to$ $\left(\pi^*\right)^2$)    & 4.63   & 4.63   & 4.71        & 4.81         & 4.70        & 4.69         & 5.73  & 4.74  \\
Glyoxal (n$^2\to$ $\left(\pi^*\right)^2$) & 4.83   & 4.97   & 5.37        & 5.56         & 5.88        & 6.56         & 6.76  & 5.54  \\
Pyrazine (n$^2\to$ $\left(\pi^*\right)^2$)& 7.35   & 7.49   & 7.90        & 8.15         & 8.43        & 8.78         & 9.17  & 8.04  \\
Tetrazine (n$^2\to$ $\left(\pi^*\right)^2$, $^1A_g$)&	3.99&	4.14&	4.55&	4.89&	5.10&	5.72&	6.18&	4.60\\
Tetrazine(n$^2\to$ $\pi^*_1\pi^*_2$, $^3B_{3g}$)&	4.77&	4.87&	5.24&	5.73&	5.92&	6.78&	7.34&	5.51\\
Tetrazine(n$^2\to$ $\pi^*_1\pi^*_2$, $^1B_{3g}$)&	5.12&	5.24&	5.62&	6.19&	6.40&	7.20&	7.60&	6.14\\
         &        &        &             &              &             &              &       &       \\
\hline
RMSE&	0.65&	0.56&	0.25&	0.18&	0.31&	0.76&	1.15&\\
RMSE (insensitive)&	0.52&	0.47&	0.21&	0.20&	0.20&	0.22&	0.71&\\
RMSE (sensitive)&	0.77&	0.64&	0.28&	0.17&	0.39&	1.05&	1.46&\\
ME&	-0.58&	-0.49&	-0.19&	0.02&	0.14&	0.54&	1.02&\\
MAX&	1.02	&0.90	&0.52&	0.33	&0.50&	1.27&	1.82&\\ \hline
\end{tabular}
\caption{Vertical excitation energies (in eV) for pure double excitations with DFT/aug-cc-pVTZ\cite{dunning1989gaussian,kendall1992electron,prascher2011gaussian}, compared to CC3 and theoretical best estimates (TBE) from Ref \citenum{loos2019reference}. CC3 values from Ref \citenum{loos2019reference} have been extrapolated to the complete basis set (CBS) limit. Root mean squared error (RMSE), mean error (ME) and maximum absolute error (MAX) relative to TBE are also reported.}
\label{tab:doubles}
\end{table}

The results shown in Table \ref{tab:doubles} demonstrate that orbital optimization with standard density functionals can achieve very good accuracy for the doubly excited states considered here, surpassing considerably more expensive wave function techniques like CC3\cite{christiansen1995response} that do not incorporate excited state orbital relaxation. Indeed, even the humble SPW92\cite{Slater,PW92} LSDA functional (that is only accurate for the uniform electron gas) has a lower root mean squared error (RMSE) than CC3! 
It appears that the meta generalized gradient approximations (mGGAs) SCAN\cite{SCAN} and B97M-V\cite{b97mv} from Rung 3 of Jacob's ladder are very accurate for the double excitations studied, yielding rather small RMSEs $\le 0.25$ eV. The PBE0\cite{pbe0} global hybrid GGA also performs well, with an RMSE of only $0.31$ eV, while the range-separated hybrid, $\omega$B97X-V\cite{wb97xv} yields rather disappointing performance in light of its good accuracy for ground state energetics\cite{mardirossian2017thirty} and properties\cite{hait2018accurate,hait2018accuratepolar}. 

The origin of this behavior could be partly understood by looking at the sensitivity of the predictions to the functional choice. The majority of the species in Table \ref{tab:doubles} show remarkably little functional sensitivity (with CH$_3$NO having a standard deviation of only 0.07 eV between predictions), but the lone pair to $\pi^*$ transitions of glyoxal, pyrazine and tetrazine show significant sensitivity to the choice of functional (standard deviation of $\ge 0.5$ eV). We therefore classify these species into a ``sensitive" subset and the remainder into a ``insensitive" one, with the subset RMSEs also reported in Table \ref{tab:doubles}. SCAN, B97M-V, PBE0 and $\omega$B97X-V give excellent (and very similar) performance for the five insensitive transitions, while LSDA and PBE perform somewhat more poorly. On the other hand, the sensitive transitions are considerably more challenging, with LSDA and PBE significantly underestimating the excitation energy, while $\omega$B97X-V significantly overestimates it. SCAN, B97M-V and PBE0 make predictions intermediate to the two extremes and consequently have low error.
These trends seem to correlate well with the delocalization error present in these functionals \cite{hait2018delocalization}, although the significant difference in performance between SPW92/PBE and SCAN/B97M-V cannot be fully explained by any delocalization based argument alone. We do however note that a similar performance gap between Rung 1-2 and Rung 3 functionals were seen for static polarizability predictions\cite{hait2018accuratepolar}, which is a global metric for accuracy of symmetry allowed singly excited states. 

Despite these limitations, all functionals tested are more accurate than the $O(N^7)$ scaling CC3 method, at only $O(N^{3-4})$ cost. The very good performance of Rung 3 functionals is encouraging in this light, as it shows that useful results can be obtained from relatively inexpensive local functionals, permitting reasonable estimate of energies of doubly excited states for very large molecular systems or even extended materials.
\begin{table}[hbt!]
\begin{tabular}{lllllllll}
\hline
 & Transition (Symmetry)   & SPW2 & PBE  & B97M-V & SCAN & PBE0 & $\omega$B97X-V & TBE  \\ \hline
\textbf{Singlet}  &              &      &      &        &      &      &         &      \\
Valence  & $n \to \pi^*$ ($^1A_2$)    & 3.81 & 3.65 & 3.84   & 3.51 & 3.62 & 3.81    & 3.97 \\
         & $\sigma \to \pi^*$ ($^1B_1$) & 8.84 & 8.58 & 8.86   & 8.40 & 8.64 & 8.83    & 9.21 \\
         & $\pi \to \pi^*$  ($^1A_1$)  & 8.72 & 8.88 & 9.59   & 9.55 & 9.78 & 9.86    & 9.26 \\
         &              &      &      &        &      &      &         &      \\
Rydberg  & $n \to 3s$ ($^1B_2$)     & 7.02 & 6.92 & 7.11   & 6.99 & 7.06 & 7.30    & 7.3  \\
         & $n \to 3p$  ($^1B_2$)   & 7.83 & 7.71 & 8.00   & 7.83 & 7.89 & 8.23    & 8.14 \\
         & $n \to 3p$  ($^1A_1$)   & 7.87 & 7.71 & 7.98   & 7.80 & 7.89 & 8.25    & 8.27 \\
         & $n \to 3p$  ($^1A_2$)   & 8.36 & 8.13 & 8.58   & 8.26 & 8.31 & 8.73    & 8.50  \\
\textbf{Triplet}  &              &      &      &        &      &      &         &      \\
Valence  & $n\to \pi^*$ ($^3A_2$)    & 3.32 & 3.29 & 3.37   & 3.12 & 3.26 & 3.45    & 3.58 \\
         & $\pi\to \pi^*$ ($^3A_1$)   & 6.58 & 6.22 & 6.02   & 5.80 & 5.84 & 6.08    & 6.07 \\
         &              &      &      &        &      &      &         &      \\
Rydberg  & $n \to 3s$  ($^3B_2$)    & 6.84 & 6.73 & 6.83   & 6.81 & 6.91 & 7.21    & 7.14 \\
         & $n \to 3p$ ($^3B_2$)    & 7.66 & 7.55 & 7.70   & 7.68 & 7.74 & 8.12    & 7.96 \\
         & $n \to 3p$ ($^3A_1$)    & 7.74 & 7.58 & 7.78   & 7.72 & 7.79 & 8.19    & 8.15 \\
\end{tabular}
\caption{ROKS singlet excitation energies and RO-$\Delta$SCF triplet excitation energies for HCHO in eV (using the aug-cc-pVTZ basis). The best theoretical estimates (TBE) has been obtained from Ref \citenum{loos2018mountaineering}. }
\label{tab:roksformal}
\end{table}

\subsection{Singly excited states of formaldehyde}
HCHO is a small molecule whose lowest lying excited states have been very well theoretically characterized\cite{loos2018mountaineering}, making it an ideal candidate for applying the SGM approach to converge ROKS for higher singlet excited states. The resulting excitation energies are shown in Table \ref{tab:roksformal}, along with $\Delta$SCF energies for the triplet state within the $M_S=1$ subspace (using RO orbitals for consistency with ROKS). Corresponding TDDFT numbers have been provided in the Supporting Information. Quantitative errors for all methods (along with corresponding values from TDDFT and some wave function theories) are reported in Table \ref{tab:formalerrors}.

\begin{table}[htb!]
\begin{tabular}{lllllll}
\hline
\textbf{Method}                & \multicolumn{2}{l}{Valence Excitations} & \multicolumn{2}{l}{Rydberg Excitations} & \multicolumn{2}{l}{All Excitations} \\\hline
\textbf{DFT protocols}         & RMSE               & ME                 & RMSE               & ME                 & RMSE             & ME               \\
SPW2                  & 0.40               & -0.16              & 0.31               & -0.30              & 0.35             & -0.24            \\
SPW92/TDDFT            & 0.73               & -0.43              & 1.26               & -1.25              & 1.07             & -0.91            \\
PBE                   & 0.39               & -0.29              & 0.45               & -0.45              & 0.43             & -0.38            \\
PBE/TDDFT             & 0.35               & -0.31              & 1.42               & -1.41              & 1.11             & -0.95            \\
B97M-V                & 0.24               & -0.08              & 0.25               & -0.21              & 0.25             & -0.16            \\
B97M-V/TDDFT          & 0.54               & -0.25              & 0.87               & -0.85              & 0.75             & -0.60            \\
SCAN                  & 0.50               & -0.34              & 0.35               & -0.34              & 0.42             & -0.34            \\
SCAN/TDDFT            & 0.49               & -0.09              & 0.65               & -0.61              & 0.59             & -0.39            \\
PBE0                  & 0.42               & -0.19              & 0.28               & -0.27              & 0.34             & -0.23            \\
PBE0/TDDFT            & 0.46               & -0.29              & 0.56               & -0.54              & 0.52             & -0.43            \\
$\omega$B97X-V               & 0.33               & -0.01              & 0.12               & 0.08               & 0.23             & 0.04             \\
$\omega$B97X-V/TDDFT         & 0.27               & -0.13              & 0.16               & -0.15              & 0.22             & -0.14            \\
\textbf{Wave function Theories}\cite{loos2018mountaineering} &                    &                    &                    &                    &                  &                  \\
CIS(D)                & 0.20               & 0.02               & 0.48               & -0.44              & 0.39             & -0.25            \\
CIS(D$_\infty$)               & 0.09               & -0.01              & 0.67               & -0.67              & 0.52             & -0.39            \\
ADC(2)                & 0.09               & -0.01              & 0.67               & -0.67              & 0.52             & -0.39            \\
CC2                   & 0.13               & 0.11               & 0.62               & -0.62              & 0.48             & -0.32            \\
CCSD                  & 0.12               & 0.04               & 0.01               & -0.01              & 0.08             & 0.01             \\
ADC(3)                & 0.24               & -0.19              & 0.35               & 0.35               & 0.31             & 0.13             \\
CC3                   & 0.03               & 0.00               & 0.04               & -0.04              & 0.04             & -0.02           
\end{tabular}
\caption{Errors (in eV) in predicting low lying excited states of HCHO (as given in Table \ref{tab:roksformal}) for various functionals, using both TDDFT (as indicated in the table) and ROKS/RO-$\Delta$SCF. Wave function theory errors have been found from values in Table S6 of Ref \citenum{loos2018mountaineering}. }
\label{tab:formalerrors}
\end{table}

The values in Table \ref{tab:formalerrors} stem from only one species, but nonetheless contain some inferences that are likely to be transferable.
First, using ROKS for singlet excited states (and RO-$\Delta$SCF for triplets) does not lead to any degradation in performance for valence excitations, consistent with previous studies \cite{kowalczyk2011assessment,kowalczyk2013excitation}. 
This is unsurprising, as standard valence excitations should not be accompanied by considerable orbital relaxation. 
Second, the situation is different for Rydberg states, where TDDFT has long been known to systematically underestimate excitation energies on account of delocalization error \cite{dreuw2005single}. ROKS/RO-$\Delta$SCF dramatically reduces errors in local functionals, often by more than a factor of 3. The residual error still stems from systematic underestimation, which is likely on account of delocalization error (which overstabilizes the diffuse density of Rydberg states relative to the ground state), 
This is similar to behavior seen for CT excited states in Ref \onlinecite{hait2016prediction}. The global hybrid PBE0 also sees a substantial reduction in error with the orbital optimized procedure, although the
range separated hybrid $\omega$B97X-V functional gives very similar behavior across both approaches. This is not too surprising, as the non-local exchange in $\omega$B97X-V guarantees correct asymptotic behavior for long-ranged particle-hole interactions (that are essential for Rydberg states) within linear response theory itself.

The overall ROKS/RO-$\Delta$SCF DFT errors for the Rydberg states compare very well with the wave function theory errors in Table \ref{tab:formalerrors}, with only the highly expensive CCSD and CC3 methods having substantially lower errors. The wave function theories however are more accurate for the valence excitations, although the  ROKS/RO-$\Delta$SCF errors are not too large for some modern functionals. Further studies involving larger datasets (like the full set presented in Ref \onlinecite{loos2018mountaineering}) and many more functionals would be necessary to determine the overall efficacy of the DFT based approaches. The relatively low errors of  ROKS/RO-$\Delta$SCF and the high computational scaling of wave function based methods however indicate considerable promise for use of ROKS/RO-$\Delta$SCF to study low-lying excited states of large systems where wave function theory is unaffordable and TDDFT unsuitable.

\section{Low lying excited states of zinc phthalocyanine}
\begin{figure}
    \centering
    \includegraphics[width=0.5\textwidth]{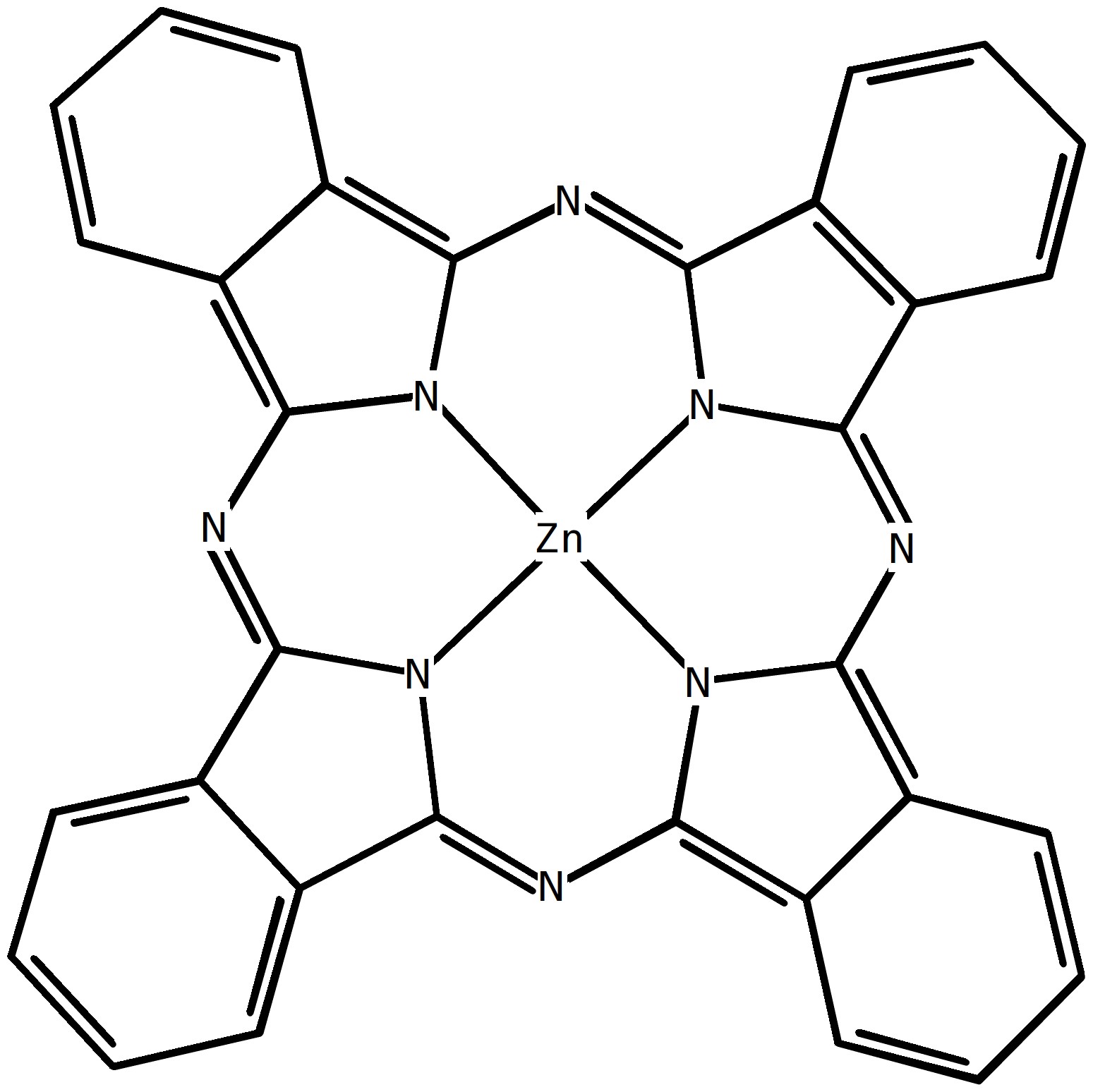}
    \caption{Zinc Phthalocyanine.}
    \label{fig:my_label}
\end{figure}
Metallophthalocyanines are species with a large, extensively $\pi$ conjugated phthalocyanine ligand coordinated to a metal atom. They share many structural features with the biologically relevant porphyrin species and possess readily tunable electronic properties that has led to use  in the electronics industry \cite{armstrong2009organic,kushto2010organic,walter2010porphyrins} and as photosensitizers\cite{bonnett1995photosensitizers}. The excited state spectra of Zn-phthalocyanine (ZnPc) has consequently been studied both experimentally\cite{edwards1970porphyrins,nyokong1987analysis,metcalf1990magnetic,mack1995band,van1989magnetic} and theoretically\cite{nguyen2001ground,ricciardi2001ground,peralta2008magnetic,theisen2015ground,wallace2017coupled}. The sheer size of this system makes DFT based approaches the reasonable choice (although a coupled cluster study with very small basis and significant virtual space truncation has been reported\cite{wallace2017coupled}) and we have consequently chosen it to demonstrate SGM's applicability to sizeable systems. We examined the symmetry allowed singlet excited states involving the twenty highest energy occupied orbitals and the ten lowest energy virtual orbitals, using the PBE0 functional and the def2-SV(P)\cite{weigend2005balanced} basis. The corresponding TDDFT excitation energies were also computed for comparison. Only the low energy (i.e. $\le 5$ eV) Q,B and N bands are reported here\cite{edwards1970porphyrins}, as high energy states have (potential) multiconfigurational character and the possibility of Rydberg like behavior\cite{wallace2017coupled} that cannot be captured without diffuse functions in the basis set. It is worth noting that ligand to metal CT (LMCT) type transitions are not possible as Zn has a full $3d^{10}$ shell. Metal to ligand CT (MLCT) transitions are possible, but appear to occur at energies $\ge 5.3$ eV with both ROKS and TDDFT. The reported transitions are therefore entirely based out of ligand orbitals. 

\begin{table}[]
\begin{tabular}{lllll}
\hline
Transition             & ROKS & TDDFT & \multicolumn{2}{l}{Experiment\cite{van1989magnetic}} \\ \hline
$2a_{1u}\mbox{(HOMO)}\to 7e_g\mbox{(LUMO)}$ & 1.97   & 2.15         & 1.89            & Q            \\
 &      &     & 2.08            & Q$^\prime$            \\
$3b_{2u}\to 7e_g\mbox{(LUMO)}$ & 3.57       & 3.56     &                 &              \\
$6a_{2u}\to 7e_g\mbox{(LUMO)}$ & 3.72   & 3.86         & 3.71            & B$_2$           \\
$2b_{1u}\to 7e_g\mbox{(LUMO)}$ & 3.96  & 3.92          &                 &              \\
$5a_{2u}\to 7e_g\mbox{(LUMO)}$ & 3.97  & 4.03          & 3.74            & B$_1$           \\
$28e_u\to 7e_g\mbox{(LUMO)}$ & 4.08    & 3.99        & 3.99            & B$_3$           \\
$2a_{1u}\mbox{(HOMO)}\to 8e_g$ & 4.13   & 4.10         &                 &              \\
$1a_{1u}\to 7e_g\mbox{(LUMO)}$ & 4.52   & 4.52         & 4.41            & N$_1$           \\
                       &              &   & 4.7             & N$_2$           \\ \hline
\end{tabular}
\caption{Symmetry allowed vertical singlet excitation energies (in eV) computed with PBE0/def2-SV(P), as compared to experimental measurements of ZnPc in Ar matrix\cite{van1989magnetic}. The experimental band assignments (Q,B etc.) have also been supplied. The TDDFT assignments have been based on the largest coefficient for transitions, in the ground state orbital basis.}
\label{tab:znpc}
\end{table}

The computed vertical excitation energies have been reported in Table \ref{tab:znpc}, along with Ar matrix experimental data \cite{van1989magnetic}. The computed ROKS and TDDFT energies agree very well with each other, showing that orbital optimization was not particularly necessary for this system. Nonetheless, the good agreement between the two approaches permits us to draw conclusions more confidently, as TDDFT possesses multiconfigurational character (within the singles subspace). The lack such multiconfigurational character therefore does not appear to affect ROKS performance here.

The lowest energy Q band for ZnPc is well separated from the rest of the spectrum, on account of the HOMO and LUMO being energetically separated from other orbitals. Our computed PBE0/def2-SV(P) ROKS energy for the Q band agrees quite well with experiment. However, we do not observe any symmetry allowed states that are close in energy to the Q$^\prime$ state that has been suggested by experimental work (via subtraction of simulated curves from observed spectra)\cite{van1989magnetic}. Previous theoretical work\cite{nguyen2001ground,ricciardi2001ground,wallace2017coupled} has also not observed such a state, indicating that it is not of electronic origin. There exists a possibility that it is a symmetry forbidden state that appears due to vibronic distortion. However, the original assignment of it being of a $A_{2u}$ state stemming from excitation of N lone pairs into $\pi^*$ levels\cite{van1989magnetic,mack1995band} is very unlikely on account of lack of any lone pairs close in energy to the HOMO (even after ignoring symmetry considerations). It is also worth noting that the HOMO$^2\to$LUMO$^2$ double excitation energy is estimated to be 3.56 eV by $\Delta$SCF with PBE0/def2-SV(P), making it an unlikely candidate for the Q$^\prime$ band. This dark state can nonetheless play a role in the photophysics/photochemistry of the system. It is, of course inaccessible to TDDFT, which illustrates a comparative strength of the $\Delta$SCF approach.

The B band is experimentally observed to be very broad, extending from approx. 3.0 eV to 4.3 eV. Ref \onlinecite{van1989magnetic} interpreted it as a combination of two transitions B$_1$ and B$_2$, although solvent phase measurements have suggested the presence of as many as five separate transitions\cite{mack1995band}. We also find 5 states with $E_u$ symmetry corresponding to that region of the spectrum, with energies roughly centered around the reported Ar matrix band maximums. Interestingly, one of those states is a transition to an unoccupied orbital that is not the LUMO (the 2$b_{1u}\to 8e_g$ excitation). It is worth noting that the B$_3$ state is distinct from the rest of the B band as it has been assigned to be a N lone pair to $\pi^*$ transition of A$_{2u}$ symmetry, and we also find a state with the same symmetry at 4.08 eV with ROKS, offering fairly reasonable agreement. 

The N band offers more of a challenge, for although we observe a state similar to the experimentally observed N$_1$ state, no state anywhere close in energy to the N$_2$ band was found with either TDDFT or ROKS. This might be a consequence of the multiconfigurational nature of the state (which could cause the ROKS energy to be too high). However, the N$_2$ band was a very weak contributor to the experimentally observed N band, and could likely have a non-electronic origin (or arise from symmetry forbidden transitions on account of vibronic perturbations). 



\section{Summary and discussion}
We have presented a general approach to converge excited state solutions for \textit{any} quantum chemistry orbital optimization technique. \textcolor{black}{A simple finite difference based implementation of the resulting Squared Gradient Minimization (SGM) approach requires only analytic orbital gradients of the energy/Lagrangian and costs approximately three times as much as standard ground state minimization (on a per iteration basis)}. SGM represents a direct minimization based alternative to the existing Maximum Overlap Method (MOM)\cite{gilbert2008self,barca2018simple}, that provides robust minimization to the stationary point closest to the initial guess at the expense of somewhat increased computational cost. It is simpler and thus more efficient, though also more initial-guess dependent, than other recently proposed excited state variational principles.\cite{ye2017sigma,shea2018communication,shea2019generalized}

Promising results were obtained within the KS-DFT framework (using the $\Delta$SCF and ROKS approaches), especially for challenging problems like charge-transfer, Rydberg and doubly excited states (using $\Delta$SCF when no electron pairs are broken and ROKS when one pair is uncoupled) that are beyond the ability of standard TDDFT to model. TDDFT nonetheless possesses the distinct advantage of being `black-box' in the sense that it permits simultaneous computation of multiple excited states without any prior knowledge about their nature/energies. TDDFT is also quite accurate for low lying valence excitations of closed-shell molecules, where state-by-state orbital optimization offers little additional benefit. It is therefore useful to list the circumstances under which usage of SGM is likely to be beneficial for applications purposes. 

SGM is the most effective when the nature of the target state can be reliably guessed, from chemical intuition or experimental data. The $Q$ band of ZnPc is a clear example of this nature, as it is quite well understood to be a HOMO$\to$LUMO type of transition. Similarly, it is also often possible to enumerate potential CT states in donor-acceptor complexes or LMCT/MLCT excitations in transition metal compounds, and directly target them. Naive enumeration of states would likely be unwise on account of a rapidly growing number of possibilities, necessitating use of narrow selection rules or extraneous information to limit the search space. 

An alternative is to first run a pilot TDDFT computation and subsequently determine which states are of CT or Rydberg nature, followed by specifically optimizing them with SGM while leaving valence excitations as is. TDDFT natural transition orbitals (NTOs) could in fact prove to be very useful initial guesses for such problems. This strategy would not be useful for double excitations (as TDDFT cannot detect them directly). Information from more sophisticated wave function approaches like CC3 could be helpful, but likely impractical due to their very high computional cost. The best way to identify potential double excitations (aside from chemical intuition) is via examination of low energy TDDFT single excitations, which might couple together. 

Further work is certainly desirable to assess the performance of ground state functionals for modeling excited states within the $\Delta$SCF and ROKS approaches. It is possible to consider extending the present approach by employing SGM to converge regularized orbital optimized MP2 (OOMP2)\cite{lee2018regularized} or even orbital optimized coupled cluster (CC) methods\cite{sherrill1998energies} for excited states. This direction potentially complements very recent work on converging CC amplitudes for excited states.\cite{zheng2019performance,lee2019excited}

From a practical standpoint, the $\Delta$SCF and ROKS ansatze constrain the current applications of SGM to single configuration excited states. This limitation can be lifted in practice by using a set of optimized excited HF determinants (i.e. abandoning DFT) as a many-electron basis for non-orthogonal Configuration Interaction (NOCI)\cite{thom2009hartree,sundstrom2014non}. NOCI-MP2\cite{yost2016size,yost2018efficient} then provides an approach to add dynamic correlation in a well defined manner and for relatively low computational cost.



\section*{Computational Details}
All calculations were performed with the Q-Chem 5.2 \cite{QCHEM4} package. 	Local exchange-correlation integrals were calculated over
a radial grid with 99 points and an angular Lebedev grid with 590 points for all atoms. 
\section*{Acknowledgment} 
	This research was supported by the Director, Office of Science, Office of Basic Energy Sciences, of the U.S. Department of Energy under Contract No. DE-AC02-05CH11231. 
\section*{Supporting Information} 
	\noindent Geometries of species studied (zip).
	
	\noindent TDDFT excitation energies and provenance of geometries (xlxs). 
\bibliography{references}
\end{document}